# On Hamiltonian approach on 3+1-dimensional space-time

K.Yu. Bliokh[1,2*]

[1]*Institute of Radio Astronomy, 4 Krasnoznamyonnaya st., Kharkov, 61002, Ukraine*
[2]*Department of Physics, Bar-Ilan University, Ramat Gan, 52900, Israel*

The paper proposes a 4-dimensional generalization of the Hamilton equations of motion to the case of the Minkowski space-time. The approach can be applied to quantum as well as to classical, non-relativistic as well as relativistic problems.

Commonly, the Hamiltonian formalism is introduced by means of a symplectic structure on the 3-dimensional coordinate space and the conjugated 3-dimensional momentum space (with possible dependence of parameters on an additional coordinate of time) (see, for instance, [1]). Such approach is accepted in non-relativistic as well as in relativistic problems, where the use of the unified 4-dimensional space-time and 4-dimensional momentum is more natural and preferable. Besides, the 4-dimensional form often simplifies the representation of problems with an electromagnetic field. Meanwhile, an immediate generalization of the Hamilton equations to 4-dimensional form is impossible – it leads to a contradiction in the energy equation (see below). In this short communication we present a simple and laconic modification of the Hamiltonian formalism for the 3+1 Minkowski space with a conjugated momentum-energy space.

Let us consider the Minkowski space-time with $(-+++)$ signature. We introduce quantum operators of contravariant 4-coordinates $r^\alpha = (r^0, \mathbf{r})$ and of covariant canonically conjugated 4-momentum $p_\alpha = (p_0, \mathbf{p})$. At that, $r^0 \equiv ct$ ($t$ is the time), $p_0 \equiv -c^{-1}E$ ($E$ is the energy operator), and the momentum operator can be represented as

$$p_\alpha = -i\hbar \partial_{r^\alpha} . \qquad (1)$$

Thus, the energy operator $E$ is not a Hamiltonian, but operator $i\hbar\partial_t$. This provides canonical commutation relations between coordinates and momentum components:

$$[r^\alpha, p_\beta] = i\hbar \delta^\alpha_\beta , \qquad (2)$$

where $\delta^\alpha_\beta$ is the Kronecker symbol.

Let us introduce now a physical system (in what follows – a particle) described by a Hamiltonian $H = H(r^\alpha, p_\alpha) \equiv H(r^\alpha, \mathbf{p})$ and a wave function $\psi$. With the quantities defined above, the Schrödinger equation (we talk about the Schrödinger equation in the wide sense, assuming that the Dirac, Maxwell, etc. equations also can be represented in such form), $i\hbar\partial_t \psi = H\psi$, takes the form

$$[H + p_0 c]\psi = 0 . \qquad (3)$$

Hence, energy operator $-p_0 c$ equals the Hamilton operator when applied to the wave function, and the Schrödinger equation (3) has the sense of a constraint imposed on the system. Let us define operator $\mathcal{H}$, which can naturally be referred to as the *modified Hamiltonian* of the system:

$$\mathcal{H}(r^\alpha, p_\alpha) = H(r^\alpha, \mathbf{p}) + p_0 c . \qquad (4)$$

With this operator the Schrödinger equation (3) takes a laconic form:

$$\mathcal{H}\psi = 0 . \qquad (3a)$$

Now we introduce the Hamilton equations of motion for the particle. In the quantum case these are the Heisenberg equations:

---
[*]E-mail: k_bliokh@mail.ru

$$\dot{\mathbf{r}} = -\frac{i}{\hbar}[\mathbf{r}, H], \quad \dot{\mathbf{p}} = -\frac{i}{\hbar}[\mathbf{p}, H], \tag{5}$$

where dot stands for the full derivative with respect to $t$. Equations (5) cannot be generalized to a 4-dimensional form immediately with usual Hamiltonian $H$, since equation $\dot{r}^0 = -\frac{i}{\hbar}[r^0, H]$ leads to a contradiction $\dot{t} = 0$. However, this can be done with the help of the modified Hamiltonian $\mathcal{H}$. Indeed, let us write down equations

$$\dot{r}^\alpha = -\frac{i}{\hbar}[r^\alpha, \mathcal{H}], \quad \dot{p}_\alpha = -\frac{i}{\hbar}[p_\alpha, \mathcal{H}]. \tag{6}$$

With Eqs. (1), (2) and (4) taken into account, space components of Eqs. (6) are equivalent to the traditional Heisenberg equations (5), whereas the time components yield

$$\dot{t} = 1, \quad \dot{p}_0 = -c^{-1}\partial_t H. \tag{7}$$

The first equation (7) is evidently true and lifts the above-mentioned contradiction, while the meaning of the second equation (7) becomes clear in the semiclassical limit.

In the semiclassical limit, when the particle has the form of a propagating wave packet (whose center corresponds to the classical particle), the Schrödinger equation (3a) and Hamilton equations (6) give rise respectively to the dispersion equation,

$$\mathcal{H}(r^\alpha, \pi_\alpha) = 0, \text{ or } \varepsilon = H(r^\alpha, \boldsymbol{\pi}), \tag{8}$$

and to the canonical Hamilton equations of classical mechanics:

$$\dot{r}^\alpha = \partial_{\pi_\alpha}\mathcal{H}, \quad \dot{\pi}_\alpha = -\partial_{r^\alpha}\mathcal{H}. \tag{9}$$

Here $\pi_\alpha = (-c^{-1}\varepsilon, \boldsymbol{\pi})$ is the 4-momentum of the corresponding classical particle ($\varepsilon$ is the energy of the classical particle); its connection with the quantum particle is realized through the de Broglie correspondence: $\pi_\alpha = \hbar k_\alpha$, where $k_\alpha = (-c^{-1}\omega, \mathbf{k})$ is the wave 4-vector of the center of the wave packet ($\omega$ is the frequency). Dispersion equation (8), as well as the Schrödinger equation (3a), plays the role of a constraint, which restricts the motion to a 7-dimensional hyper-surface in the 8-dimensional phase space $(r^\alpha, \pi_\alpha)$ (see [2]). Space components of Eqs. (9) represent classical Hamilton equations with the usual Hamiltonian $H$, the time (zero) component of the first Eq. (9) coincides with the first Eq. (7), and the zero component of the second Eq. (9) results in

$$\dot{\varepsilon} = \partial_t H. \tag{10}$$

This equation is well-known in classical mechanics (see, for instance, [3]); it represents the energy variation law in a non-conservative system. Thus, we have a good ground to regard the quantum equation (7) as true and similar to Eq. (10) in the meaning. In particular, in the case of the conservative system, i.e. when the Hamiltonian is independent from $t$, Eqs. (7) and (10) result in the energy conservation law:

$$\dot{p}_0 = \dot{\varepsilon} = 0. \tag{11}$$

The proposed approach is naturally generalized to the case when an external gauge field is present, that has been used in papers [5]. At that, quantum equations (6) conserve their form, while semiclassical Eqs. (9) acquire additional field terms due to non-trivial commutators of variables in the presence of a gauge field. For instance, for a particle with charge $e$ in an external electromagnetic field with tensor $F_{\alpha\beta}$, $[p_\alpha, p_\beta] = i\hbar\frac{e}{c}F_{\alpha\beta}$, and semiclassical motion equations read

$$\dot{r}^\alpha = \partial_{\pi_\alpha}\mathcal{H}, \quad \dot{\pi}_\alpha = -\partial_{r^\alpha}\mathcal{H} + \frac{e}{c}F_{\alpha\beta}\dot{r}^\beta. \tag{12}$$

In summary, we have constructed a laconic Hamiltonian formalism on the 3+1-dimensional space-time. This approach can be used in non-relativistic as well in relativistic, quantum as well as classical problems. The principal step, which enables one to reveal a natural 4-dimensional generalization of the Hamilton equations of motion, is the introduction of the modified Hamiltonian (4) that includes an explicit dependence on the zero component of the momentum operator (energy). In such formalism the energy operator is not the Hamiltonian but the operator of partial differentiation with respect to time. These operators coincide only on the wave function of the particle, and this constraint is nothing else but the Schrödinger equation (3a). It is worth noting that an analogous Hamiltonian formalism with zero-valued Hamiltonian (8) is commonly used in the geometrical optics of smoothly inhomogeneous media (see, for instance, [4]).

I am grateful to S.-C. Zhang for drawing my attention to Ref. 2. The work was partially supported by INTAS (Grant No. 03-55-1921) and by Ukrainian President's Grant for Young Scientists GP/F8/51.